\documentclass[final,5p,times,twocolumn]{elsarticle}

\usepackage{graphicx}
\usepackage{epsfig}

\usepackage{amssymb}
\usepackage{amsthm}

\usepackage{lineno}
\usepackage{hyperref}
\usepackage{color}
\hypersetup{colorlinks=blue}

\biboptions{sort&compress}

\journal{arXiv}

\begin{document}
\begin{frontmatter}



\title{Signatures of room-temperature superconductivity emerging
in two-dimensional domains within the new $\rm{Bi/Pb}$-based ceramic
cuprate superconductors at ambient pressure}


\author{S. Dzhumanov$^1$\corref{cor1}}
\cortext[cor1]{Corresponding author. Fax:(8-10-99871) 289 36 65}
\ead{dzhumanov@inp.uz}

\author{D.D. Gulamova$^2$\corref{}}
\author{Sh.S. Djumanov$^3$\corref{}}

\address{$^1$Institute of Nuclear Physics, Uzbek Academy of Sciences, 100214, Ulugbek,
Tashkent, Uzbekistan}
\address{$^2$Institute of Material Science,
Uzbekistan Physics-Sun Scientific Production Association, Uzbek
Academy of Sciences, Tashkent, 100084, Uzbekistan}

\address{$^3$Tashkent State Technical University, Tashkent 100095, Uzbekistan}

\begin{abstract}
We predict the possibility of realizing room-temperature
superconductivity in different two-dimensional (2D) domains within
the ceramic high-$T_c$ cuprate superconductors at ambient pressure
and experimentally confirm this prediction of 2D room-temperature
superconductivity in the newly derived $\rm{Bi/Pb}$-based ceramic
cuprate superconductors containing many grain boundaries, interfaces
and multiplate blocks. We argue that, in these high-$T_c$ materials,
besides bulk superconductivity in three-dimensional (3D) domains
there is also strongly enhanced 2D superconductivity emerging in the
3D-2D crossover region well above the superconducting transition
temperature $T_c$. We study the possibility of the existence of
distinct 3D and 2D superconducting phases in high-$T_c$ ceramic
cuprates, in which the unconventional (tightly-bound polaronic)
Cooper pairs behave like bosons and condense below certain critical
temperatures into 3D and 2D Bose superfluids in 3D and 2D domains.
We show that the superconducting transition temperature in 2D
domains is much higher than in 3D domains and can reach up to room
temperature. We report signatures of room-temperature
superconductivity occurring at different grain boundaries and 3D/2D
interfaces and in multiplate blocks within the ceramic
superconductors $\rm{Bi_{1.7}Pb_{0.3}Sr_{2}Ca_{n-1}Cu_{n}O_{y}}$
(where $n=2-30$), synthesized by using the new melt technology in a
large solar furnace (in Parkent). The samples of these materials
synthesized under the influence of concentrated solar energy have
the bulk $T_c$ values ranging from 100 K  to about 140 K and the
more higher superconducting transition temperatures, possibly even
as high as room temperature in the 3D-2D crossover region. The
remnant 2D superconductivity in newly derived $\rm{Bi/Pb}$-based
ceramic cuprate superconductors is observed at temperatures
$T\gtrsim200-300$ K well above the bulk $T_c$ and the onset of
room-temperature superconductivity is evidenced by the observations
of a sharp step-like drop in the resistance and a well-detectable
partial Meissner effect at around 300 K and ambient pressure.
\end{abstract}


\end{frontmatter}


\section{Introduction}\label{sec:level1}

Remarkable progress in condensed matter physics is often driven by
discoveries of new superconducting materials and superfluid (Fermi
or Bose) liquids. In particular, the discovery of $\rm{La}$-based
high-$T_c$ cuprate superconductors \cite{1,2,3} and then the
dramatic increase of the critical temperature $T_c$ of the
superconducting transition in other discovered $\rm{Y}$-, $\rm{Bi}$-
and $\rm{Hg}$-based cuprate superconductors to above 90 K \cite{4},
110 K \cite{5} and 130 \cite{6} ushered in a new era of physics and
technology. These important discoveries stimulated efforts to find
new materials with even higher superconducting transition
temperatures, possibly even close to room temperature. Various
experiments showed \cite{3,7,8} that the superconductivity is
occurred in the bulk three-dimensional (3D) cuprate material rather
than in the $\rm{CuO_2}$ plane. Therefore, the above values of $T_c$
observed in different high-$T_c$ cuprates correspond to the bulk
superconducting transition temperatures.

Starting from 1993, there has been considerable interest in the
increase of $T_c$ in some families of cuprate superconductors to a
maximum value at different applied pressures \cite{9,10,11} and the
record values of $T_c\simeq153-164$ K reported for $\rm{Hg}$-based
cuprate superconductors are remained still unchanged. For a long
time, despite intense research effors (see
Refs.\cite{12,13,14,15,16}), the increasing of $T_c$ up to room
temperature in various high-$T_c$ cuprates was remained a difficult
problem. So far, room-temperature superconductivity was not reported
for high-$T_c$ cuprate materials even under pressure.

Recently, there has been growing interest in the discovery of
room-temperature superconductivity in other classes of materials and
the new so-called high-$T_c$ hydrides have been synthesized at very
high pressures \cite{17,18,19}. However, room-temperature
superconductors discovered under high pressures will not have a
wide-range practical applications. Therefore, in the last decade,
new attempts were made to synthesize the promising ceramic cuprate
superconductors with the onset temperature of the superconducting
transition, $T_c^{onset}\gtrsim150-180$ K at ambient (atmospheric)
pressure, by using more advanced technology \cite{20,21}. Prominent
among high-$T_c$ cuprate superconductors are the so-called
$\rm{Bi/Pb}$-based ceramic cuprate compounds, such as
$\rm{Bi_{1.7}Pb_{0.3}Sr_{2}Ca_{n-1}Cu_{n}O_{y}}$ (with $n=2-30$),
which may present an alternative path to realizing room-temperature
superconductivity at atmospheric pressure.

For many decades, there has been speculation (see, e.g.,
Refs.\cite{14,18}) that Little's model of high-$T_c$
superconductivity in one-dimensional organic polymers with
polarizable side chains \cite{22} and Ginzbur's model of
two-dimensional (2D) alternating conducting/insulating sandwich
layers \cite{23} may be possible routes to room-temperature
superconductivity. Such early theoretical predictions are based on
the Bardeen-Cooper-Schrieffer (BCS)-like theory of Fermi-liquid
superconductivity. Still, most researchers trying to find the new
high-$T_c$ superconductors make predictions on the basis of the
BCS-like and Migdal-Eliashberg theories of Fermi-liquid
superconductivity regarding the possibility of increasing the
critical temperature $T_c$ up to room temperature (see
Refs.\cite{15,16,17}). However, the high-$T_c$ cuprates undergoing a
$\lambda$- superconducting transition at $T_c$ \cite{24,25} just
like $\lambda$-transition in liquid $^4$He can be in the regime of
Bose-liquid superconductivity \cite{26,27} and the BCS-like theories
are incapable of predicting the actual $T_c$ in such unconventional
superconductors.

In this work, we present theoretical and experimental results of
room-temperature superconductivity emerging in different 2D domains
within the newly derived $\rm{Bi/Pb}$-based ceramic cuprate
superconductors containing many grain boundaries, interfaces and
mutiplate blocks. We show that, in these ceramic high-$T_c$
materials, besides bulk 3D superconductivity there is also strongly
enhanced 2D superconductivity occurring well above the bulk $T_c$ in
the 3D-2D crossover region. We argue that the unconventional Cooper
pairs in high-$T_c$ cuprates behave like bosons and condense below
certain critical temperatures into 3D and 2D Bose superfluids in 3D
and 2D domains. We examine the possibility of the existence of 3D
and 2D superconducting phases in ceramic cuprate superconductors
predicted by the theory of 3D and 2D Bose superfluids. We find that
superconducting transition temperature in 2D domains is much higher
than that in 3D domains and the highest critical temperature $T_c$
for some high-$T_c$ cuprates, e.g., $\rm{Bi/Pb}$-based ceramic
cuprate superconductors can reach up to room temperature in 2D
domains. We report experimental signatures of room-temperature
superconductivity emerging at different grain boundaries and
interfaces and in multiplate blocks within the ceramic cuprate
superconductors $\rm{Bi_{1.7}Pb_{0.3}Sr_{2}Ca_{n-1}Cu_{n}O_{y}}$
(with $n=2-30$), synthesized by using the new melt technology under
the influence of concentrated solar energy. We claim that the
occurrence of such a room-temperature superconductivity in these
high-$T_c$ materials is evidenced by the observations of a sharp
drop in the resistance and a pronounced partial Meissner effect at
around 300 K and ambient pressure.

\section{Theoretical blackground}

The cuprate compounds are typical polar materials and
charge-transfer-type Mott-Hubbard insulators \cite{28}. Therefore,
in these materials the charge carriers (holes or electrons)
introduced by doping in a polar crystal strongly interact with
optical phonons \cite{29} and their self-trapping is favorable just
like the self-trapping of holes in ionic crystal of alkali halides
\cite{30,31}. In doped high-$T_c$ cuprates, the Cooper pairing of
self-trapped carriers (polarons) results in the formation of
tightly-bound (bosonic) Cooer pairs \cite{32}. These unconventional
Cooper pairs in 3D high-$T_c$ cuprates condense into a 3D Bose
superfluid below the bulk $T_c$ \cite{32,33}. We now consider the
ceramic cuprate superconductors consisting of 3D and 2D domains.
Then the actual $T_c$ in 3D domains is determined from the
self-consisting solutions of the integral equations of a 3D Bose
superfluid \cite{34}

\begin{eqnarray}\label{Eq.1}
\frac{1}{\gamma_B}=\int^{\xi_{BA}}_0\sqrt{\varepsilon/\xi_{BA}}\frac{coth\left[\sqrt{(\varepsilon+\tilde{\mu}_B)^2-\Delta^2_B}/2k_BT\right]}{\sqrt{(\varepsilon+\tilde{\mu}_B)^2-\Delta^2_B}}d\varepsilon,
\end{eqnarray}
\begin{eqnarray}\label{Eq.2}
\frac{2\rho_B}{D_B}=\int^{\infty}_0
\sqrt{\varepsilon}\times\nonumber\\\times
\left\{\frac{\varepsilon+\tilde{\mu}_B}{\sqrt{(\varepsilon+\tilde{\mu}_B)^2-\Delta^2_B}}coth\left[\frac{\sqrt{(\varepsilon+\tilde{\mu}_B)^2-\Delta^2_B}}{2k_BT}\right]-1
\right\}d\varepsilon,
\end{eqnarray}
where $\gamma_B=D_B\tilde{V}_B\sqrt{\xi_{BA}}$ is the interboson
coupling constant in a 3D Bose superfluid, $\rho_B$ is the density
of attracting (superfluid) bosons,
$D_B=m_B^{3/2}/\sqrt{2}\pi^2\hbar^3$ is the density of states, $m_B$
is the mass of free bosons, $\tilde{\mu}_B$ is the renormalized
chemical potential of an interacting Bose gas,
$\tilde{V}_B=V_{BA}-V_{BR}/[1+D_{B}V_{BR}(\sqrt{\xi_{BR}}-\sqrt{\xi_{BA}})]$
is the effective interaction potential between bosons, $\xi_{BA}$
and $\xi_{BR}$ are the cutoff parameters for the attractive $V_{BA}$
and repulsive $V_{BR}$ parts of the interboson interaction
potential, $\Delta_B$ is the coherence parameter (i.e. superfluid
order parameter) of condensed bosons.

The critical temperature $T_c^{2D}$ of the superconducting
transition in 2D domains is determined from the self-consistent
solutions of the integral equations of a 2D Bose superfluid
\cite{34}

\begin{eqnarray}\label{Eq.1}
\frac{1}{\gamma_B}=\int^{\xi_{BA}}_0\frac{coth\left[\sqrt{(\varepsilon+\tilde{\mu}_B)^2-\Delta^2_B}/2k_BT\right]}{\sqrt{(\varepsilon+\tilde{\mu}_B)^2-\Delta^2_B}}d\varepsilon,
\end{eqnarray}
\begin{eqnarray}\label{Eq.2}
\frac{2\rho_B}{D_B}=\int^{\infty}_0\times\nonumber\\\times
\left\{\frac{\varepsilon+\tilde{\mu}_B}{\sqrt{(\varepsilon+\tilde{\mu}_B)^2-\Delta^2_B}}coth\left[\frac{\sqrt{(\varepsilon+\tilde{\mu}_B)^2-\Delta^2_B}}{2k_BT}\right]-1\right\}d\varepsilon,
\end{eqnarray}
where $\gamma_B=D_B\tilde{V}_B$ is the interboson coupling constant
in a 2D Bose superfluid, $\rho_B$ is the density of 2D superfluid
bosons, $D_B=m_B/2\pi\hbar^2$.

Numerical solutions of Eqs. (1) and (2) determining the bulk
$T_c=T^{3D}_c$ in 3D superconducting domains can be obtained for
arbitrary $\gamma_B$. While the approximate analytical solutions of
these equations are obtained for $\gamma_B<1$ \cite{34}. In
particular, the bulk $T_c$ at $\gamma_B>0.3$ can be determined
approximately from the expression

\begin{eqnarray}\label{Eq.1}
T_c=T^*_{BEC}[1+c\gamma_B\sqrt{\sqrt{2}k_BT^*_{BEC}/\xi_{BA}}],
\end{eqnarray}
where $T^*_{BEC}=3.31\hbar^2\rho^{2/3}_B/k_Bm^*_B$ is the
renormalized Bose-Einstein condensation temperature in a 3D
Bose-liquid, in which $m_B$ is replaced by the renormalized mass
$m^*_B$ of interacting bosons defined similarly in Ref.\cite{35},
$c=\pi^{3/2}/3.918$.

The solutions of Eqs. (3) and (4) determining the critical
temperature $T^{2D}_c$ in 2D superconducting domains are obtained
analytically \cite{34} and the superconducting transition
temperature at grain boundaries and interfaces and in multiplate
blocks within the ceramic high-$T_c$ cuprate materials for arbitrary
$\gamma_B$ is determined from the following relation:

\begin{eqnarray}\label{Eq.1}
T_c^{2D}=-\frac{T_0^*}{ln[1-exp(-2\gamma_B/(\gamma_B+2))]},
\end{eqnarray}
where $T^*_0=2\pi\hbar^2\rho_B/k_Bm^*_B$.

We use the analytical expressions (5) and (6) to examine the
possibility of the existence of distinct 3D and 2D superconducting
phases in ceramic high-$T_c$ materials and to predict the existence
of two distinctly different regimes of high-$T_c$ superconductivity
in them and the possible route to room-temperature
superconductivity. In order to determine the characteristic
superconducting transition temperatures in the bulk and the 3D-2D
crossover region of high-$T_c$ cuprates, we estimate $T_c$ and
$T^{2D}_c$ by assuming that the mass $m_p$ of polaronic carriers in
3D domains is of the order of $2m_e$ \cite{36} (where $m_e$ is the
mass of free electrons) and in 2D domains is somewhat larger than
$2m_e$. It is natural to assume that, in high-$T_c$ cuprates, the
preformed bosonic (polaronic) Cooper pairs have the mass $m_B=2m_p$
and only the interacting bosons with the renormalized mass
$m^*_B>m_B$ can condense into Bose superfluids at $T\lesssim T_c$
(in 3D domains) and $T\lesssim T^{2D}_c$ (in 2D domains). By taking
$m_p\simeq2m_e$, $m_B=2m_p$, $m^*_B=1.05m_B$ and
$\rho_B\simeq4\cdot10^{19}$ $cm^{-3}$ for 3D superconducting domains
in ceramic cuprate materials, we find $T^*_{BEC}\simeq81.7$ K. Using
Eq.(5) we further estimate the bulk $T_c$ by assuming that two
bosonic Cooper pairs interact with each other by means of exchange
optical phonons having relatively low energy
$\hbar\omega_0\simeq0.03$ eV and the cutoff energy $\xi_{BA}$ for
the attractive part of the interboson interaction potential can be
replaced by $\hbar\omega_0$. Then, at $\gamma_B=0.7$ the analytical
expression (5) for the bulk $T_c$ predicts the value of
$T_c\simeq1.508T^*_{BEC}\simeq123$ K. We can now estimate the
superconducting transition temperature in 2D domains by assuming
that $m_p\simeq3m_e$, $m_B=2m_p$, $m^*_B=1.05m_B$ and
$\rho_B\simeq2\cdot10^{13}$ $cm^{-2}$. Under these assumptions, we
obtain $T^*_0\simeq176$ K. If we take the same value of
$\gamma_B=0.7$ for a 2D Bose superfluid in the analytical expression
(6), we find $T^{2D}_c\simeq1.105T^*_0\simeq195$ K. The above
predictions of the theory of Bose-liquid superconductivity in
high-$T_c$ cuprates indicate that the expected value of $T^{2D}_c$
in the 3D-2D crossover region will be much higher than the expected
value of $T_c$ in the bulk. Therefore, the 3D domains in ceramic
high-$T_c$ cuprates become non-superconducting above the bulk
$T_c(=T^{3D}_c)$, but high-$T_c$ superconductivity is still
maintained in 2D domains in a wide temperature range above $T_c$.
Such a remnant 2D superconductivity persisting at grain boundaries
and interfaces and in multiplate blocks within the ceramic
high-$T_c$ cuprate materials might be encouraging in achieving
room-temperature superconductivity at atmospheric pressure. In order
to make this argument more quantitative, the results of our
numerical calculations of the critical superconducting transition
temperature $T^{2D}_c$ as a function of $\gamma_B$ are presented in
Fig.1. As may be seen in Fig.1, at some values of $\rho_B$ and
$m_B$, the critical temperature $T^{2D}_c$ strongly depending on
$\gamma_B$ eventually reaches to room-temperature at a certain value
of $\gamma_B\gtrsim0.8$.

\begin{figure}[!htp]
\begin{center}
\includegraphics[width=0.45\textwidth]{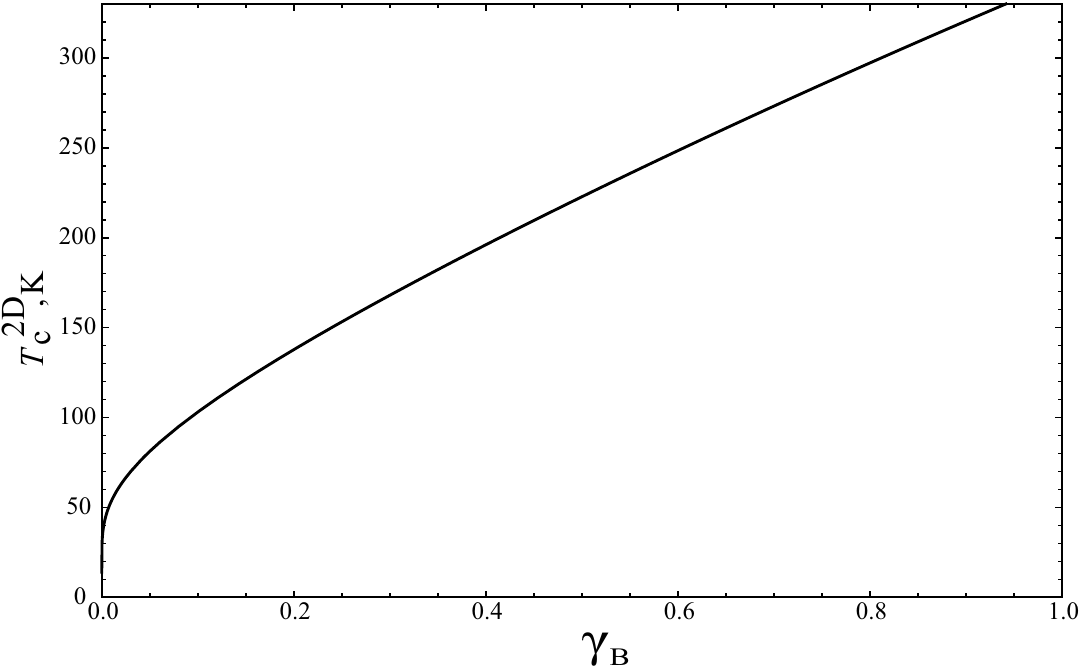}
\caption{\label{Fig1} The calculated critical temperature $T^{2D}_c$
of the superconducting transition in 2D systems as a function of
interboson coupling constant $\gamma_B$. The parameters used are
$\rho_B\simeq2.8\cdot10^{13} cm^{-2}$, $m_p=3m_e$, $m_B=2m_p$ and
$m^*_B\simeq1.05m_B$.}
\end{center}
\end{figure}

\begin{figure}[!htp]
\begin{center}
\includegraphics[width=0.45\textwidth]{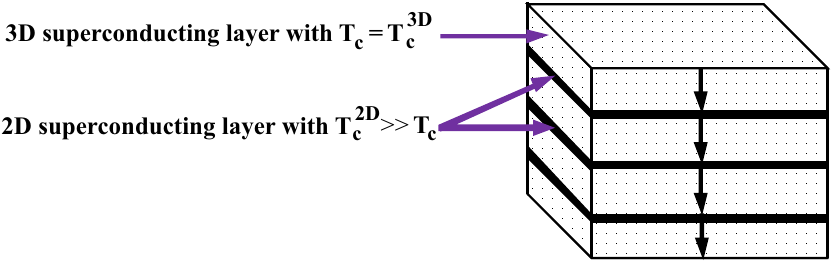}
\caption{\label{Fig2} Alternating 3D non-superconducting/2D
superconducting sandwich layers in ceramic cuprate superconductors
above the bulk $T_c$. Bold-type arrows indicate a crossover from 3D
non-superconducting (metallic) state to 2D superconducting state in
these sytems.}
\end{center}
\end{figure}

Here we point out a possibility that many grain boundaries parallel
to each other, interfaces and parallel multiplate blocks in
specially grown ceramic cuprate superconductors (see, e.g.,
experimental results presented in Sec.3) will be involved in
alternating 3D non-superconducting/2D superconducting sandwich
layers above the bulk $T_c$. As can be seen in Fig.2, the crossover
from the bulk 3D regime to surface 2D regime of superconductivity in
these alternating 3D/2D sandwich layers might be possible path to
realizing room-temperature superconductivity in such distinctive
ceramic high-$T_c$ materials. Apparently, there is some experimental
evidence for the signatures of superconducting transition well above
the bulk $T_c$ in different families of high-$T_c$ cuprates. In
particular, experimental results on the anomalous resistive
transitions between 125 and 260 K in the cuprate superconductors
$\rm{Y_{3}Ba_{4}Cu_{7}O_{x}}$ (with the bulk $T_c\simeq90$ K)
\cite{12} indicate that superconductivity above the bulk $T_c$
emerges at grain boundaries in accordance with the prediction of the
above expression (6) for $T^{2D}_c$. Further, there is also
experimental evidence for the signatures of superconductivity at
temperatures much higher than the bulk $T_c$ in other high-$T_c$
cuprates, where resistive transitions were also observed at a
temperature of about 260 K \cite{13}. Concurrently, the possibility
of traces of superconductivity at temperatures $T>>T_c$ in the
systems $\rm{HgBa_{2}Ca_{n-1}Cu_{n}O_{2n+2+\delta}}$ and
$\rm{Bi_{2}Sr_{2}CaCu_{2}O_{8}}$ was reported \cite{13}. Thus, the
above key experimental findings provided the confirmation of the
predictions of the theory of 3D and 2D Bose superfluids regarding
the possibility of superconductivity well above the bulk $T_c$ in
$\rm{Y}$-, $\rm{Bi}$-and $\rm{Hg}$-based cuprate superconductors.
Among these high-$T_c$ materials, some $\rm{Bi}$-based ceramic
cuprate superconductors may be the promising systems to investigate
the 3D-2D crossover regime of superconductivity, which might be
relevant to room-temperature superconductivity. In the following, we
report experimental evidence of 2D room-temperature
superconductivity occurring in the 3D-2D crossover region within the
newly derived $\rm{Bi/Pb}$-based ceramic cuprate superconductors,
which contain many grain boundaries, interfaces and multiplate
blocks.

\section{New melt technology for synthesizing of $\rm{Bi/Pb}$-based ceramic cuprate superconductors under the influence of solar energy}

The possibility of producing superconducting materials with the
highest critical temperature to a considerable extent depends on the
methods of their synthesizing. The solid-state reaction method is
often used to synthesize the samples of high-$T_c$ cuprate
superconductors. However, this method has the following
shortcomings: the synthesis of samples is the multi-stage process
and the preparation of well-textured samples is difficult. In this
regard, the melt technologies of sample synthesis are promising in
obtaining high-$T_c$ cuprate superconductors with higher critical
temperature and have certain advantages over other method of
synthesizing high-$T_c$ superconducting samples. For the synthesis
of the samples of high-$T_c$ cuprate superconductors, the resistance
and induction furnaces, optical furnace, laser and other energy
sources were used for melting a mixture of starting materials. In
the last years, the new melt technology for synthesizing
$\rm{Bi/Pb}$-based ceramic cuprate superconductors were developed by
using the concentrated solar energy for melting a mixture of source
materials \cite{37}. The advantages of this method in the synthesis
of the samples of high-$T_c$ ceramic materials have been described
elsewhere \cite{20,21}. The new melting technology based on the use
of solar energy allows us to obtain the well-textured ceramic
cuprate superconductors
$\rm{Bi_{1.7}Pb_{0.3}Sr_{2}Ca_{n-1}Cu_{n}O_{y}}$ (with $n=2-30$), in
which the onset of superconductivity occurs at critical temperatures
close to room temperature. The samples of these ceramic cuprate
superconductors were prepared by a melt process in a large solar
furnace (in Parkent).

\begin{figure}[!htp]
\begin{center}
\includegraphics[width=0.4\textwidth]{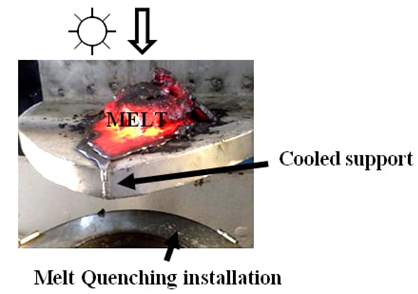}
\caption{\label{Fig3} Melting process of a mixture of source
materials and synthesis of ceramic cuprate superconductors
$Bi_{1.7}Pb_{0.3}Sr_2Ca_{n-1}Cu_nO_y$ in a large solar furnace.}
\end{center}
\end{figure}

\begin{figure}[!htp]
\begin{center}
\includegraphics[width=0.25\textwidth]{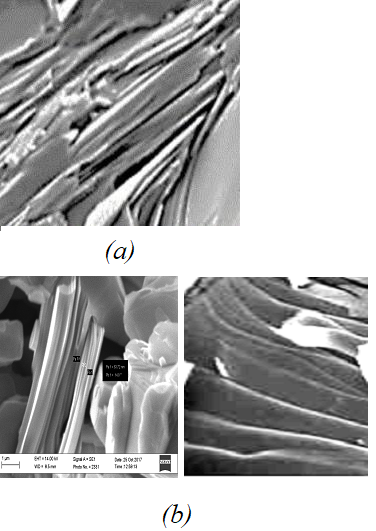}
\caption{\label{Fig4} Microstructures of ceramic cuprate
superconductors (a) $Bi/Pb$ (2.2. 29.30 ) and (b)  $Bi/Pb$
(2.2.19.20) \cite{21} obtained by using the electron microscopy and
the microscopy Solver Next and SEM.}
\end{center}
\end{figure}

Nominal compositions,
$\rm{Bi_{1.7}Pb_{0.3}Sr_{2}Ca_{n-1}Cu_{n}O_{y}}$ were prepared from
adequate mixtures of precursors $\rm{Bi_{2}O_3}$, $\rm{PbO}$,
$\rm{SrSO_3}$, $\rm{CaO}$ and $\rm{CuO}$. Mixtures of these starting
source materials were then put in a solar furnace and melted by
concentrated solar radiation with a radiant flux of about 420-480
$W/cm^2$ in this furnace, which is shown in Fig.3. The appropriate
temperature gradient in the melt process is of order 1480 $^0$C. The
samples were prepared in the form of bars or plates with the size
$5\times5\times45 mm$ and in the form of disks with a diameter of
$14-26 mm$ and a thickness of $2.3-5.4 mm$. The microstructure of
the prepared samples were studied using the microscopy solver Next,
SEM and electron microscopy Zeiss. Phase analysis of the samples was
performed by $X$-ray diffraction (DRON UM-1) technique (using
$\rm{CuK_\alpha}$ radiation and $\rm{Ni}$ filler). Samples of the
new $\rm{Bi/Pb}$-based cuprate superconductors contain many grain
boundaries parallel to each other, twin boundaries, interfaces and
multiplate blocks (see Fig.4).

\section{Experimental evidence of room-temperature superconductivity in $\rm{Bi/Pb}$-based ceramic cuprate superconductors}

According to the $X$-ray diffraction data and observed resistive
transitions, the synthesized samples of $\rm{Bi/Pb}$-based cuprate
superconductors are multiphase materials, where different
superconducting phases are formed at different critical
temperatures. In particular, 3D superconducting phases with
different bulk $T_c$ values are observed in the temperature range
$100 K\lesssim T_c\lesssim 130 K$. Analyses of the microstructure of
ceramic samples using electron microscopies showed that the new
superconducting systems
$\rm{Bi_{1.7}Pb_{0.3}Sr_{2}Ca_{n-1}Cu_{n}O_{y}}$ consist of many
small and large grains (i.e. nanocrystals), twin domains and closely
packed parallel multiplate blocks, as shown in Fig.4. The
microstructure of the chip of ceramic samples obtained from
glass-crystalline plates and annealed at temperatures
$T\simeq843-850$ $^0$C for 3-94 h represent closely packed
multilamellar blocks, which contain coupled stacks of many quasi-2D
parallel layers with nano-sized thickness.

\begin{figure}[!htp]
\begin{center}
\includegraphics[width=0.45\textwidth]{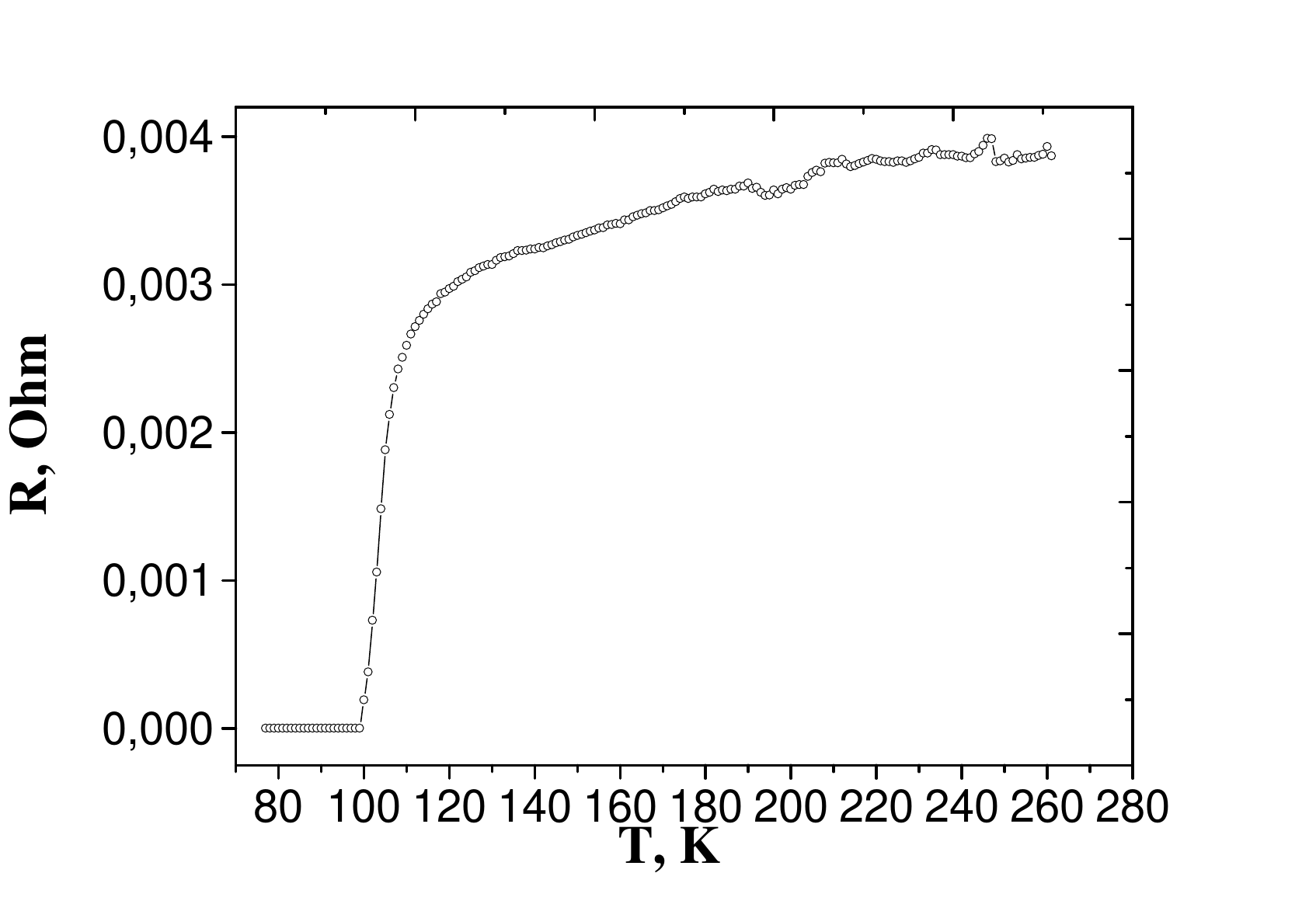}
\caption{\label{Fig5} Temperature dependence of the resistance for
the sample of $\rm{Bi/Pb}-2245$.}
\end{center}
\end{figure}

The samples of $\rm{Bi/Pb}$-based ceramic cuprate superconductors
were used to study the superconducting properties in a wide
temperature range from 77 to 300 K. The critical temperatures of the
superconducting transitions were determined by the
resistance-temperature measurements carried out using the standard
four-probe technique with silver past contacts. The contact
resistance was far less than $1\Omega$ in each case. The
superconducting transition temperature in a series of the studied
samples is defined by the maximum slope of the resistive transition
in the resistance-temperature curves. Fig.5 shows the
resistance-temperature dependence of the synthesized sample of
$\rm{Bi_{1.7}Pb_{0.3}Sr_{2}Ca_{n-1}Cu_{n}O_{y}}$
($\rm{Bi/Pb}-2245$), where the onset of superconductivity in 2D
domains seemingly occurs at the temperature $T^{onset}_c\simeq210$ K
well above the bulk $T_c\simeq120$ K. In Fig.6, we show the
resistance-temperature dependence for the synthesized sample of
$\rm{Bi/Pb}-2267$. For this sample, the onset of superconductivity
is observed at room temperature $T^{onset}_c\simeq395$ K and a sharp
step-like drop in the resistance is observed at around this
temperature. The onset of superconductivity in the studied sample of
$\rm{Bi_{1.7}Pb_{0.3}Sr_{2}Ca_{n-1}Cu_{n}O_{y}}$
($\rm{Bi/Pb}-221920$)occurs at $T^{onset}_c\simeq310$ and a sharp
drop of the resistance is observed at around this temperature (see
Fig.7). Fig.8 shows the resistance-temperature curve for the
synthesized sample of
$\rm{Bi_{1.7}Pb_{0.3}Sr_{2}Ca_{n-1}Cu_{n}O_{y}}$
($\rm{Bi/Pb}-222930$). First, the resistance of this sample
increases slightly from 350 K down to 295 K and then a sharp drop in
the resistance is observed around the temperature
$T^{onset}_c\simeq295$. The superconducting transition exhibits a
step at about 295 K.

Upon lowering the temperature, other superconducting transitions in
the studied samples of $\rm{Bi/Pb}-2245$, $\rm{Bi/Pb}-2267$,
$\rm{Bi/Pb}-221920$ and $\rm{Bi/Pb}-222930$ are observed at
$T_c\simeq100-130$ K. Our experimental results indicate that the
bulk 3D superconductivity in these $\rm{Bi/Pb}$-based ceramic
cuprate materials occurs only at more lower temperature than the
onset temperature of 2D superconductivity emerging at different
grain boundaries and 3D/2D interfaces and in multiplate blocks. The
superconducting transition in the multiphase superconductors
$\rm{Bi_{1.7}Pb_{0.3}Sr_{2}Ca_{n-1}Cu_{n}O_{y}}$ occurs first at
about room temperature and is manifested as a step-like resistive
transition in the resistance-temperature curve. Such a resistive
transition is indicative of the presence of more than one
superconducting phase and the persistence of remnant 2D
superconducting phase in these ceramic materials above the bulk
$T_c$ up to room temperature. Here we notify that the 3D and 2D
superconducting phases coexist below the bulk $T_c$ and the ceramic
samples of the $\rm{Bi/Pb}$-based cuprate superconductors below this
critical temperature predominantly consist of 3D superconducting
phase.

\begin{figure}[!htp]
\begin{center}
\includegraphics[width=0.45\textwidth]{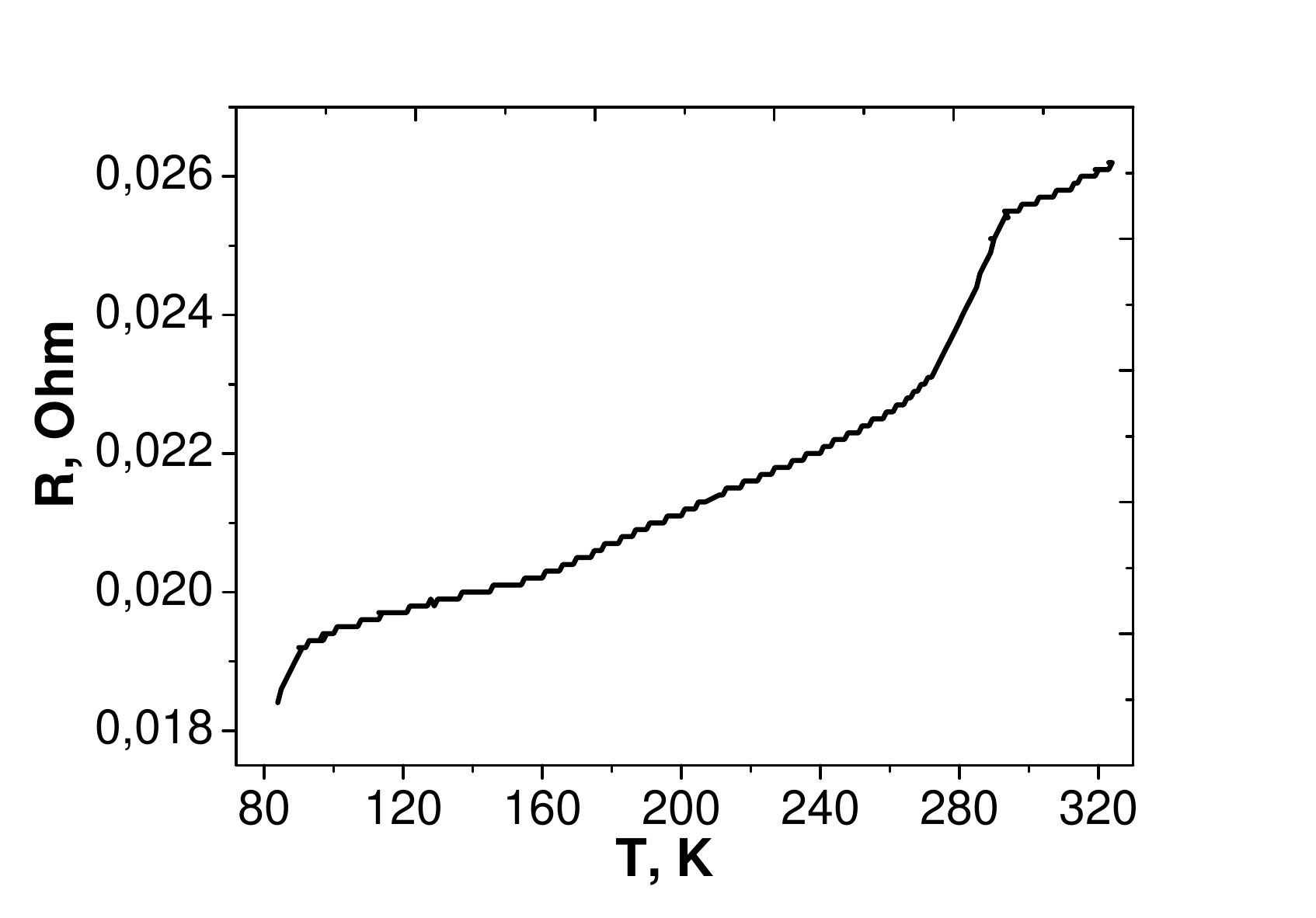}
\caption{\label{Fig6} The resistance-temperature dependence for the
sample of $\rm{Bi/Pb}-2267$ .}
\end{center}
\end{figure}

\begin{figure}[!htp]
\begin{center}
\includegraphics[width=0.35\textwidth]{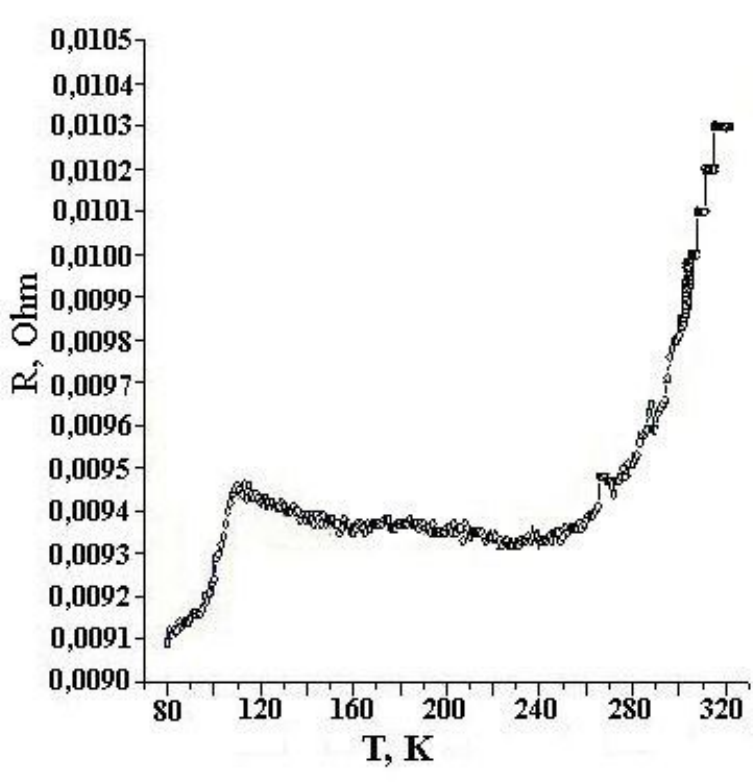}
\caption{\label{Fig7} The resistance-temperature dependence for the
sample of $\rm{Bi/Pb}-221920$.}
\end{center}
\end{figure}

Another key criterion for emerging of room-temperature
superconductivity in the synthesized samples of
$\rm{Bi_{1.7}Pb_{0.3}Sr_{2}Ca_{n-1}Cu_{n}O_{y}}$ is the detection of
the presence of a pronounced Meissner effect below a characteristic
temperature of the step-like resistive transition occurring at
around 300 K. We detected an incomplete Meissner effect in the most
samples of these ceramic superconducting materials at room
temperature.

In Fig.9, we illustrate the detection of such a well-defined
Meissner effect, that is, the expulsion of magnetic flux from the
$\rm{Bi/Pb}$-based ceramic cuprate superconductors at room
temperature. Upon lowering the temperature, fractional Meissner
effect in these materials increases gradually and remained always
smaller than 100\% even below the bulk $T_c$, very likely due to the
presence of the insulating phase in underdoped and even in optimally
doped high-$T_c$ cuprates \cite{38,39,40}. In insulating domains,
the Cooper pairs do not exist in $k$-space any longer and become
localized pairs (i.e. bipolarons) in real space \cite{41}, as
observed in $\rm{Bi}-2212$ \cite{42}. The variations of the Meissner
effect in ceramic high-$T_c$ cuprate superconductors reflect
variations of the insulating volume and superconducting volumes of
3D and 2D domains. One can assume that a partial superconducting
volume of 2D domains in the studied samples of $\rm{Bi/Pb}$-based
cuprate superconductors and related partial Meissner effect detected
at room temperature varies from 5\% to 25\%.

\begin{figure}[!htp]
\includegraphics[width=0.45\textwidth]{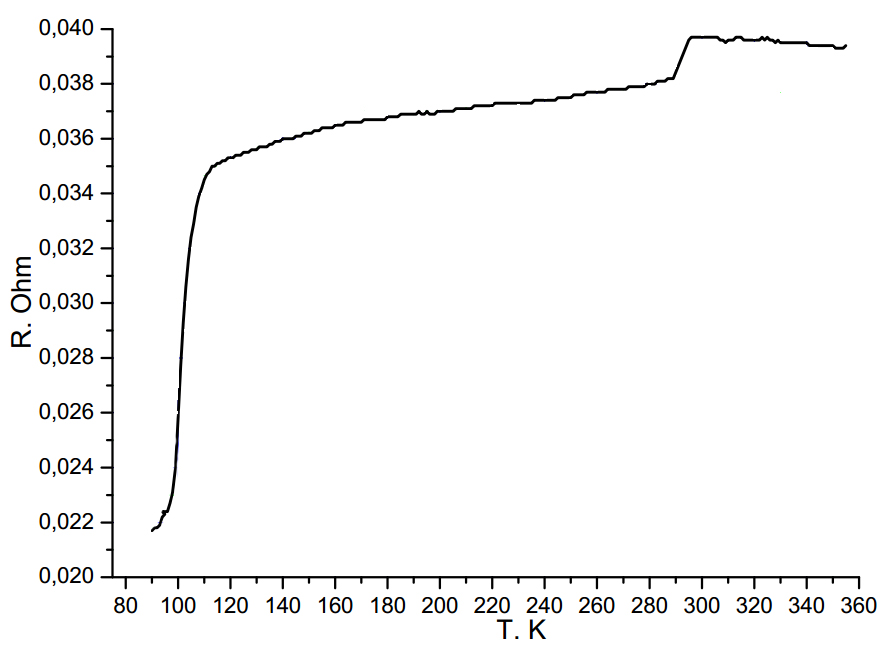}
\caption{\label{Fig8} The resistance-temperature dependence for the
sample of $\rm{Bi/Pb}-222930$.}
\end{figure}

\begin{figure}[!htp]
\begin{center}
\includegraphics[width=0.5\textwidth]{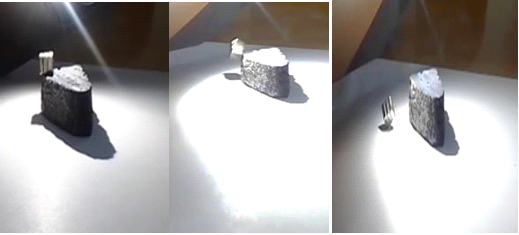}
\caption{\label{Fig9} The partial Meissner effect detected in the
$\rm{Bi/Pb}$-based ceramic cuprate superconductor
$\rm{Bi_{1.7}Pb_{0.3}Sr_{2}Ca_{n-1}Cu_{n}O_{y}}$ (n=30) at room
temperature.}
\end{center}
\end{figure}

\section{Conclusion}

We have presented the theoretical and experimental results
concerning the possibility of the existence of 3D and 2D
superconducting states in specially synthesized multiphase ceramic
cuprate superconductors
$\rm{Bi_{1.7}Pb_{0.3}Sr_{2}Ca_{n-1}Cu_{n}O_{y}}$ (with $n=2-30$).
These newly derived high-$T_c$ materials consist of 3D and 2D
superconducting domains and are particularly interesting from the
viewpoint of novel superconducting properties and room-temperature
superconductivity of such systems. We have argued that the
high-$T_c$ cuprates are unconventional (bosonic) superconductors, in
which the tightly-bound (poloronic) Cooper pairs behave like bosons
and condense into 3D and 2D Bose superfluids in 3D and 2D domains.
We have shown that, in ceramic high-$T_c$ cuprate superconductors,
besides bulk 3D superconductivity there is also strongly enhanced 2D
superconductivity emerging in the 3D-2D crossover region and
persisting well above the bulk $T_c$. We have examined the
possibility of the existence of distinctly different 3D and 2D
superconducting phases in these materials predicted by the theory of
Bose-liquid superconductivity. We have determined the critical
superconducting transition temperatures in 3D and 2D systems, which
depend on the effective mass of interacting bosons, the density of
superfluid bosons and the interboson coupling constant. We have
found that the superconducting transition temperatures in bosonic
cuprate superconductors is much higher in 2D domains than in 3D
domains and the bulk superconductivity is destroyed above
$T_c(=T^{3D}_c)$, but 2D superconductivity persists up to the onset
temperature $T^{onset}_c=T^{2D}_c>>T^{3D}_c$. We have predicted the
possibility of realizing room-temperature superconductivity at
different grain boundaries and 3D/2D interfaces and in multiplate
blocks within the new ceramic cuprate superconductors. Apparently,
some families of such cuprate superconductors synthesized by using
advanced technologies might be particularly promising in the
experimental search for room-temperature superconductivity at
ambient pressure. We have used the new technology for synthesizing
the ceramic cuprate superconductors with very high critical
temperatures, ever observed previously, and finally discovered
room-temperature superconductivity in the $\rm{Bi/Pb}$-based ceramic
high-$T_c$ materials
$\rm{Bi_{1.7}Pb_{0.3}Sr_{2}Ca_{n-1}Cu_{n}O_{y}}$, synthesized by
using the new melt technology in a large solar furnace, with the
highest critical temperatures $T^{onset}_c\simeq295-310$ K at
ambient pressure. The samples of $\rm{Bi/Pb}$-based cuprate
superconductors contain grain boundaries parallel to each other,
twin boundaries and interfaces and multilamellar blocks represening
the alternating 3D non-superconducting/2D superconducting sandwich
layers above the bulk $T_c$, which are favorable for realizing
room-temperature superconductivity. The remnant 2D superconductivity
persisting at different grain boundaries and 3D/2D interfaces and in
multilamellar blocks up to high temperatures
$T=T^{onset}_c\simeq295-310$ K and the onset of room-temperature
superconductivity in the synthesized samples of $\rm{Bi/Pb}$-based
ceramic cuprate materials are evidenced by the observations of a
sharp step-like drop in the resistance and a pronounced partial
Meissner effect at around 300 K. Thus, by the use of the series of
$\rm{Bi/Pb}$-based ceramic cuprate materials, we have succeeded in
finding the new class of room-temperature superconductors  at
ambient pressure.

\section*{Acknowledgments}
We would first like to thank B.S. Yuldashev for his attention and
encouragement in the process of carrying-out this work. We would
also like to thank our colleagues I. Khidirov, B.L. Oksengendler,
U.T. Kurbanov and E.Kh. Karimbaev for valuable discussions and
suggestions. This work was support by the Foundation of the
Fundamental Research, Grant No $\Phi$-$\Phi$A-2021-433.

\end{document}